\documentclass[12pt]{article}

\renewcommand{\sectionmark}[1]{}
\renewcommand{\subsectionmark}[1]{}
\usepackage[utf8]{inputenc}
\usepackage[top=1in, bottom=1.25in, left=1.25in, right=1in]{geometry}
\usepackage{algorithm}
\usepackage{algpseudocode}
\usepackage{natbib}
\usepackage{amsmath}
\usepackage{amssymb}
\usepackage{hyperref}
\usepackage{bookt abs}
\usepackage{longtable}
\usepackage[table]{xcolor}
\usepackage{bm}
\usepackage{pdflscape}
\usepackage{tabu}
\usepackage{subcaption}
\usepackage{setspace}
\usepackage{needspace}
\hypersetup{colorlinks=true,linkcolor=blue,filecolor=blue,citecolor=blue}

\usepackage{tikz}
\usepackage[labelfont=bf]{caption}
\usepackage{relsize}
\usepackage{xifthen}
\usepackage{titlesec}
\titlespacing\section{0pt}{\parskip}{\parskip}

\allowdisplaybreaks

\numberwithin{intassumption}{assumption}

\usepackage{parskip}
\expandafter\def\expandafter\normalsize\expandafter{%
    \normalsize%
    \setlength\abovedisplayskip{5pt}%
    \setlength\belowdisplayskip{5pt}%
    \setlength\abovedisplayshortskip{-8pt}%
    \setlength\belowdisplayshortskip{2pt}%
}

\title{BreakLoops: A New Feature for the Multi-Gene, Multi-Cancer Family History-Based Model, Fam3Pro}
\author{\fontsize{12pt}{12pt}\selectfont Nicolas Kubista$^{1,2,\dagger}$, Ryan Hernandez-Cancela$^{1,2,\dagger}$, Jianfeng Ke$^{1,2}$, Romain Berquet$^{3}$, \\
\fontsize{12pt}{12pt}\selectfont Gavin Lee$^{1,2}$, Giovanni Parmigiani$^{1,2}$, Danielle Braun$^{1,2,*}$}
\date{\fontsize{12pt}{16pt}\selectfont $^1$Department of Biostatistics, Harvard T.H. Chan School of Public Health \\
$^2$Department of Data Science, Dana Farber Cancer Institute \\
$^3$École Polytechnique Fédérale de Lausanne (EPFL)
\\[0.75em]
\today
\\[0.75em]
\fontsize{10pt}{10pt}\selectfont
$^\dagger$These authors contributed equally to this work.\\
*\textit{Email:} dbraun@mail.harvard.edu
}

\begin{document}

\maketitle
\begin{abstract}
Previously, we presented PanelPRO, now known as Fam3PRO, an open-source R package for multi-gene, multi-cancer risk modeling with pedigree data. The initial release could not handle pedigrees that contained cyclic structures called loops, which occur when relatives mate. Here, we present a graph-based function called \texttt{breakloops} that can detect and break loops in any pedigree. The core algorithm identifies the optimal set of loop breakers when individuals in a loop have exactly one parental mating, and extends to handle cases where individuals have multiple parental matings. The algorithm transforms complex pedigrees by strategically creating clones of key individuals to disrupt cycles while minimizing computational complexity. Our extensive testing demonstrates that this new feature can handle a wide variety of pedigree structures. The \texttt{breakloops} function is available in Fam3Pro version 2.0.0. This advancement enables Fam3Pro to assess cancer risk in a wider range of family structures, enhancing its applicability in clinical settings.
\end{abstract}

\section{Introduction}
Fam3Pro (previously known as PanelPRO) is an R package for multi-syndrome, multi-gene risk modeling for individuals with a family history of cancer \citep{lee2021multi}. The Fam3Pro package is used to estimate the probabilities of carrying germline pathogenic variants, as well as the future risk of developing associated cancers, utilizing family history information from a pedigree data frame. Fam3Pro predicts carrier probabilities leveraging the “peeling and paring” algorithm, which is based on the “peeling” algorithm originally introduced by Elston and Stewart \citep{elston1971general}. 

\begin{figure}
    \centering
    \includegraphics[width=1\linewidth]{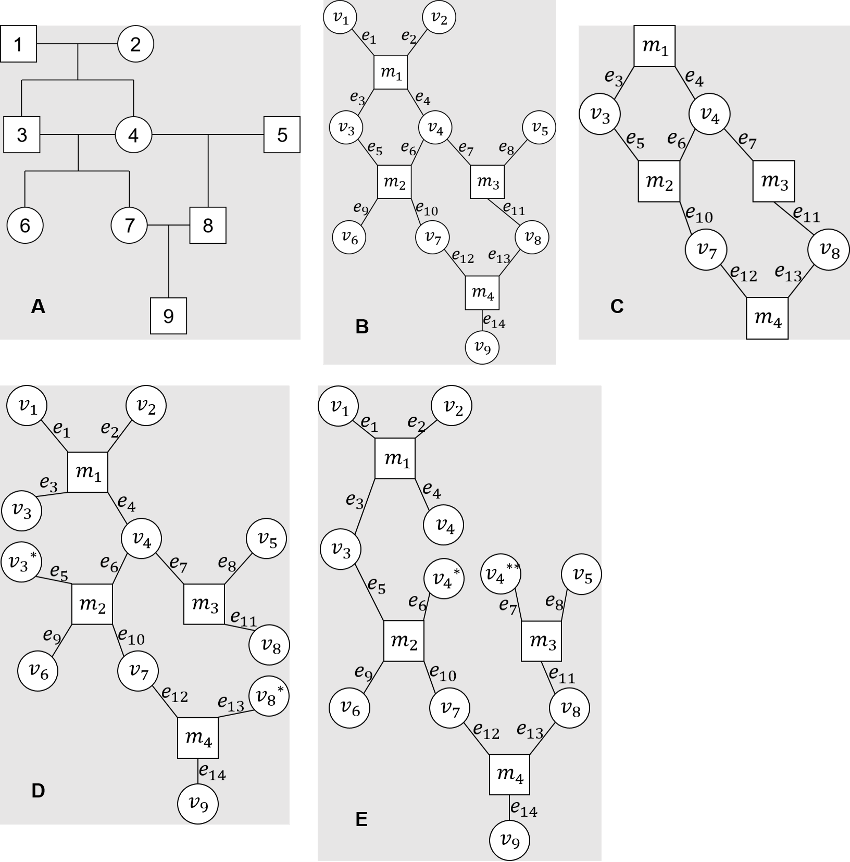}
    \caption{A pedigree example with two loops and two different loop breaking approaches (panels D and E) . Individual 4 mates with their sibling 3, and individuals 7 and 8, who are half-siblings, also mate. (A) The initial pedigree. (B) The pedigree represented as an undirected graph. (C) The trimmed graph after removing all the leaves. (D) The original graph with a clone of ID 3 and a clone of ID 8. (E) The original graph with two clones of ID 4.}
    \label{fig:1}
\end{figure}

A key input to the Fam3Pro function is the pedigree data. Some pedigrees may contain loops or cyclic paths that occur when relatives mate, known as consanguinity \citep{lange1975extensions}. For example, a mating between two siblings would create a pedigree loop. For example, in Figure \ref{fig:1}A) we show a pedigree with two loops. In this case, a loop, or a path that starts and ends at the same individual can be traced if we begin at ID 3, follow the connecting line to ID 4, travel up towards both parents (ID 1 and 2), and descend down the leftmost line back to ID 3. Notice that the only person who is visited twice on this path is ID 3, the person we started with, and the path forms a visible circle in the pedigree chart creating a loop. The pedigree contains a second loop that can be traced by starting at ID 7, traversing to ID 8, then following the connection to ID 4, moving upward to the parents (ID 1 and 2), descending to ID 3, continuing to ID 6, and finally returning to ID 7. This second loop occurs due to a mating between half-siblings, as ID 7 and ID 8 share one common parent (ID 4). Intuitively, pedigree loops can be identified by searching for these circular structures.

The “peeling and paring” algorithm mentioned above was not designed to handle pedigree loops because the conditional independence assumption of genotypes breaks down in cyclical structures. These loops create exponentially growing inheritance path combinations, as each cycle introduces interdependent genotype possibilities that must be jointly resolved. This combinatorial explosion renders exact calculations computationally intractable \citep{fernando1993efficient}. One remedy involves “breaking” pedigree loops by creating clones of individuals \citep{lange2002}. Cloning involves creating a copy of an individual with genetic and phenotypic information identical to the original individual, and we call the original individual a loop breaker. Clones replace loop breakers in their parental roles. By disconnecting these loop breakers from their downstream relatives, the pedigree loops are broken, allowing the algorithm to proceed error-free.

Assuming each loop breaker is only cloned once, the complexity of computing the carrier probability after loop-breaking increases $|G| = |G_1| \times \cdots \times |G_t|$ fold, where $t$ is the number of loop breakers and $|G_i|$ is the total number of possible genotypes of the $i$-th loop breaker. Here, a genotype represents a binary vector where each index represents a germline pathogenic variant (e.g., BRCA1), and each binary entry indicates whether an individual carries said pathogenic variant (yes/no). Many algorithms try to minimize this increase in complexity by finding an optimal set of loop breakers: that is, the set which has the minimal number of $|G| = \prod_i |G_i|$, or equivalently, $\log|G| = \sum_i \log|G_i|$. When every individual involved in a loop has exactly one mating where they are a parent, the problem of finding the set with minimal $|G|$ can be transformed into the classical Maximum Spanning Tree (MST) problem using graph theory \citep{kruskal1956shortest, prim1957shortest}. Thus, an optimal solution can be found by applying well-known algorithms like Prim’s or Kruskal’s algorithm. If at least one person involved in a loop has two or more matings where they are a parent, no algorithm is guaranteed to be optimal, but the greedy algorithm discussed by Becker and Geiger \citep{becker1998automatic} can find a close-to-optimal solution when only a few loops are present.

In this article, we introduce a new feature for Fam3Pro to handle pedigrees with loops called \texttt{breakloops}. The core of the new feature is a new graph-based algorithm called \texttt{breakloops} that can toggle between Prim’s algorithm and the greedy algorithm depending on the situation. We conducted extensive testing to ensure \texttt{breakloops} can break a wide variety of loops. Furthermore, we verified that the feature does not significantly affect the runtime of Fam3Pro.

\section{Methods}
\label{sec:meth}
\subsection{Input Pedigrees and Undirected Graphs}

In the context of this work, a pedigree is a diagram showing the relationships between family members (the proband and their relatives) as well as their cancer status. Pedigree information is stored in data frames, and the pedigree structure is defined by three columns: \textit{ID}, \textit{MotherID}, and \textit{FatherID}. Figure~\ref{fig:1}A provides an example of a pedigree. In this example, there are two loops in the pedigree: one loop is created by \textit{ID} 3 mating with his sister \textit{ID} 4, resulting in offspring \textit{ID} 6 and 7; the other loop is formed by \textit{ID} 7 mating with her half-brother \textit{ID} 8, leading to offspring \textit{ID} 9. This example represents an extraordinary case chosen for illustration purposes to demonstrate our approach in a complex family structure with multiple interconnected loops. While such consanguineous matings creating multiple loops within a single pedigree are uncommon in clinical practice, this example effectively showcases how our method systematically handles even the most intricate pedigree structures with multiple interrelated loops.

To utilize existing graph methodologies to break these loops, we must transform the pedigree into an undirected connected graph. A graph \(G\) is a data structure comprising a set of vertices \(V\) and a set of edges \(E\). Each edge connects two vertices, and an edge \(e_{a,b}\) can be represented as \((v_a, v_b)\), where \(v_a\) and \(v_b\) are connected vertices. A graph \(G\) is deemed \textit{undirected} if each edge lacks a directional component, and \textit{connected} if, \(\forall u,v \in V\), there exists a path between any \(u\) and \(v\) in the graph. A path of length \(n\) from vertex \(v_0\) to vertex \(v_n\) in \(G\) is a sequence of \(n\) distinct edges: \(\langle e_1, e_2, \dots, e_n \rangle\) such that \(e_i = (v_{i-1}, v_i)\) when \(1 \leq i \leq n\). Details on how to construct an undirected and connected graph \(G\) corresponding to a pedigree are given in Definition 1.

\textbf{Definition 1.} A pedigree \(P\) can be represented as an undirected and connected graph \(G = (V, E)\).
\label{def:1}

First, a set of vertices, \(V_1\), which represents the set of different individuals in \(P\), is constructed. Hence, every individual \(i \in P\) is associated with a corresponding vertex \(v_i \in V_1\). For a pedigree \(P\) consisting of \(n\) individuals, \(V_1\) contains \(n\) vertices.
Subsequently, a second set of vertices, \(V_2\), which represents the set of different matings in \(P\), is created. For each pair of individuals \(i\) and \(j\) who are known to have mated and had offspring (i.e. they are a mating pair), we create a vertex \(m_{i,j}\). Thus, \(V_2 = \bigcup m_{i,j}\) \(\forall i,j \in P\) such that \(i\) and \(j\) mated and have offspring in the pedigree. We assign mates their own vertices because edges can only connect two vertices, and we cannot show that an individual is linked to their partner and their child using the same edge.
Lastly, the set of edges \(E\) is created. For each mating \(m_{i,j}\) in \(V_2\), we have an edge between \(v_i\) and \(m_{i,j}\) and an edge between \(v_j\) and \(m_{i,j}\). Additionally, for the \(k\)-th child of the mating pair \(i\) and \(j\) (\(c_{i,j,k}\)), we have an edge between the child and \(m_{i,j}\). Hence, individuals \(i\) and \(j\) are connected to mating \(m_{i,j}\) as parents, and each child \(c_{i,j,k}\) is connected to \(m_{i,j}\) as offspring. For brevity, we shall use \(m_i\) to represent each mating instead of \(m_{i,j}\), and when we mention "graph", we mean "undirected, connected graph".

Furthermore, an important property of \(G\) is that each cycle in \(G\) represents a loop in the original pedigree. For reference, a cycle is a path where only the first and last vertices are the same. Proposition 1 elaborates on this relationship.

\textbf{Proposition 1.} \(P\) has a loop \(\iff\) \(G\) has a cycle, where \(G\) is the graph associated with \(P\).\\

\textit{Proof.} \(\Rightarrow\) Assume that \(P\) is a pedigree with a loop. By definition, we can find a first path, \(p_1\), in \(P\) from individual \(i\) to individual \(j\) and a second path, \(p_2\), from individual \(j\) to individual \(i\), such that \(p_1\) and \(p_2\) have distinct nodes except for the endpoints. Now, if we consider the corresponding graph \(G\), \(p_1\) will yield a corresponding path \(p'_1\), and \(p_2\) will yield a path \(p'_2\). By construction, we know that both paths \(p'_1\) and \(p'_2\) have distinct edges associated with distinct vertices except for the first and last vertices.

If we consider the path \(p = \langle e_{0,i}, \dots, e_{ik,n}, e_{n,jk}, e_{j,0}\rangle\), which is the union of paths \(p'_1\) and \(p'_2\), we have a set of distinct edges where the first and last vertices are identical. By definition, \(p\) is a cycle, so \(G\) is a cyclic graph.

\(\Leftarrow\) Conversely, assume \(G\) is a cyclic graph. By definition, \(G\) contains a cycle, which corresponds to a loop in the original pedigree \(P\). Therefore, if \(G\) has a cycle, \(P\) must have a loop.

The corresponding graph for the pedigree in Fig.~\ref{fig:1}A is given in Fig.~\ref{fig:1}B. There are two cycles in Fig.~\ref{fig:1}B, \(\langle m_1 - e_4 - v_4 - e_6 - m_2 - e_5 - v_3 - e_3 - m_1 \rangle\), and \(\langle v_4 - e_7 - m_3 - e_{11} - v_8 - e_{13} - m_4 - e_{12} - v_7 - e_{10} - m_2 - e_6 - v_4 \rangle\), each of which corresponds to a loop in the pedigree in Fig.~\ref{fig:1}A.

To better visualize these cycles, we can remove extraneous vertices and edges to produce a trimmed graph. This entails recursively removing or "trimming" leaves, which are vertices connected to only one edge. The remaining vertices represent the individuals (and matings) involved in the loops. Fig.~\ref{fig:1}C shows the graph after trimming the leaves from Fig.~\ref{fig:1}B.

\subsection{Detecting Loops}

In the previous implementation of the Fam3Pro package (v 1.0.0) without the \texttt{breakloops} function, pedigree loop detection is managed by the checkMating function. This function traverses the pedigree to ensure that no individual is involved in a mating that would create a loop. However, the checkMating function has limitations in identifying complex loops and may fail to detect intricate relationships that form such loops. For example, in the case where a proband's granddaughter marries a proband's nephew, the checkMating function may overlook this loop, potentially resulting in an infinite loop error during the computation.

To address these limitations, we propose a more robust and comprehensive method for loop detection based on graph theory. We begin by considering a graph as a \textit{tree} if it is both connected and acyclic. We define an individual as a \textit{founder} if neither of their parents is included in the pedigree. We define an individual as an \textit{offspring} if both of their parents are known in the pedigree. Let \(n_m\) and \(n_0\) denote the number of different matings and offspring in the pedigree, respectively. Each mating contributes two edges to the associated graph \(G\) since both parents are connected to that mating vertex. Each offspring contributes one edge to \(G\) since each child is connected to a mating vertex. This implies that the number of edges in the graph associated with a pedigree is equal to twice the number of matings plus the number of offspring in the pedigree, i.e.,

\begin{equation}
\label{eq:edges}
|E| = 2 \times n_m + n_0
\end{equation}

Let \(n_i\) denote the number of individuals in the pedigree. The number of vertices in the graph \(G\) is simply the number of individuals plus the number of matings in the pedigree, i.e.,

\begin{equation}
\label{eq:vertices}
|V| = n_i + n_m
\end{equation}

\cite{bondy1976graph} established a fundamental property of trees in graph theory, which has significant implications for detecting cycles in graphs and loops in pedigrees:

\textbf{Theorem 1 (Bondy and Murty).} If \(G = (V, E)\) is a tree, then \(|E| = |V| - 1\).
\label{theorem1}\\

This theorem provides a simple yet powerful criterion for identifying cycles in graphs. Specifically, a graph \(G\) is a tree if and only if it contains no cycles, as it must satisfy the condition \(|E| = |V| - 1\). Since the graph associated with a pedigree \(G\) is always connected, the presence of cycles can be determined by comparing the number of edges to the number of vertices.

\cite{bondy1976graph} also proved a more general result for connected graphs:

\textbf{Corollary 1 (Bondy and Murty).} If \(G\) is a connected graph, then \(|E| \geq |V| - 1\).
\label{col1}\\

This corollary establishes a lower bound on the number of edges in any connected graph. Equality in this relation characterizes trees, while strict inequality implies the presence of at least one cycle.

By substituting Equations~\ref{eq:edges} and~\ref{eq:vertices} into Theorem \ref{theorem1} and Corollary \ref{col1}, we can derive a simple formula to detect loops in a pedigree: if \(n_m + n_0 = n_i - 1\), then the pedigree contains no loops; however, if \(n_m + n_0 > n_i - 1\), then the pedigree contains at least one loop. Thus, constructing the associated graph \(G\) is unnecessary to detect cycles; these counts from the original pedigree suffice. 

The pseudo-code for the loop-detection algorithm, \texttt{Checkloops}, is shown in Algorithm \ref{algo1}.

\begin{algorithm}[H]
\caption{\texttt{Checkloops}: Check if there are loops in the pedigree}
\begin{algorithmic}[1]
\Require A pedigree \(P\)
\Ensure Return 1 if there is at least one loop; 0 if no loop

\State \(n_m \gets\) number of different matings
\State \(n_0 \gets\) number of offspring
\State \(n_i \gets\) number of individuals in the pedigree

\If{\(n_m + n_0 > n_i - 1\)}
    \State \Return 1 \Comment{There is at least one loop}
\Else
    \State \Return 0 \Comment{No loops}
\EndIf
\label{algo1}
\end{algorithmic}
\end{algorithm}

\subsection{Selecting Loop Breakers}

One can arbitrarily break a loop in the pedigree by randomly selecting individuals in the trimmed graph as loop breakers, severing their ties to downstream relatives, and filling the gaps with their corresponding clones. For example, a possible set of loop breakers for Fig.~\ref{fig:1}C is \textit{ID 3} and \textit{ID 8}. The modified pedigree after adding clones of \textit{ID 3} and \textit{ID 8} (\(v_3^*\), \(v_8^*\)) is given in Fig.~\ref{fig:1}D. However, a more efficient way to break the loops is to select \textit{ID 4} as the only loop breaker and add two clones of \textit{ID 4} to the pedigree, as shown in Fig.~\ref{fig:1}E.

\begin{figure}
    \centering
    \includegraphics[width=1\linewidth]{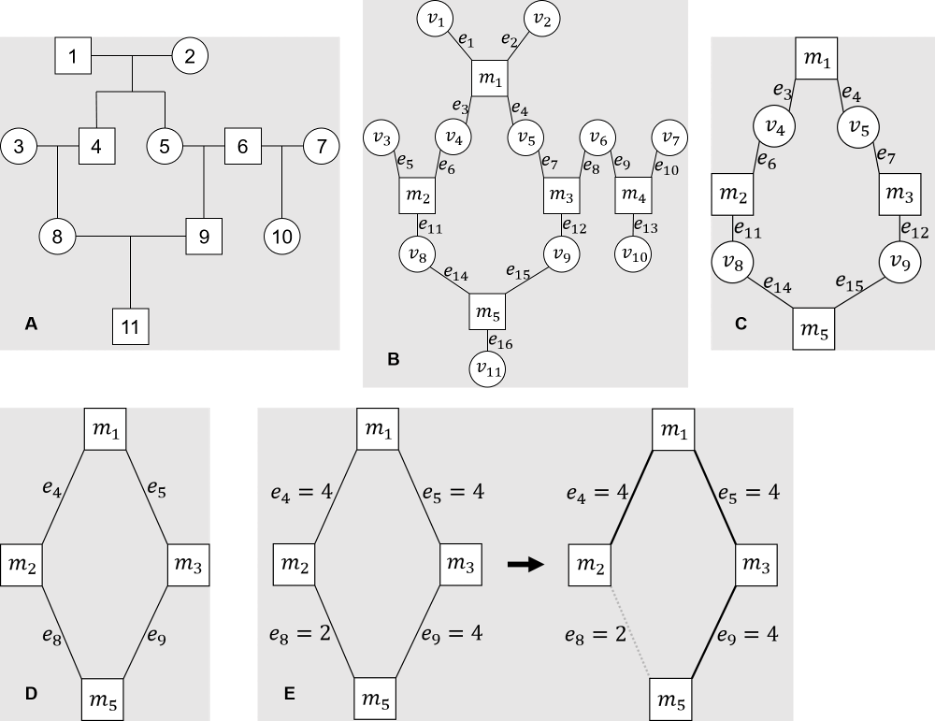}
    \caption{A pedigree example without individuals having multiple matings in the loop. (A) The initial pedigree. (B) The undirected graph representation. (C) The trimmed graph after recursively removing all the leaves. (D) The sub-graph of the trimmed graph. (E) The MST of the sub-graph when given an example set of weights, which indicates that individual 8 is the ideal loop breaker in this example}
    \label{fig:2}
\end{figure}

When every individual in the trimmed graph (i.e., everyone involved in at least one loop) is connected to exactly one mating in the trimmed graph where they are a parent, we refer to this as a "no-multiple-matings" case. Since each person is also connected to a mating where they are an offspring, everyone is connected to exactly two matings. In other words, we can say that everyone has a degree equal to two in the trimmed graph, where "degree" represents the number of edges connected to an individual. A pedigree example without multiple matings in the loops, and its graph representation, are given in Fig.~\ref{fig:2}A and Fig.~\ref{fig:2}B, respectively. Note that even though \textit{ID 6} has two matings, his mating with \textit{ID 7} (\(m_4\)) is not inside the loop. As the leaves are recursively trimmed from the graph, \textit{ID 10}, \textit{ID 7}, \(m_4\), and \textit{ID 6} are eventually removed. The trimmed graph of Fig.~\ref{fig:2}B is shown in Fig.~\ref{fig:2}C.

We can now construct a sub-graph \(G'\) based on the trimmed graph. The vertices in \(G'\) correspond only to the mating vertices of the trimmed graph of \(G\). Every path of the form \(m_k - v_i - m_j\) is replaced with an edge \(e_i\) which directly connects the mating vertices \(m_k\) and \(m_j\). The original edges that connected \(m_k\) and \(v_i\) and \(m_j\) and \(v_i\) are deleted. The individual vertex \(v_i\) is also removed. An example of the sub-graph \(G'\) based on the trimmed graph in Fig.~\ref{fig:2}C is given in Fig.~\ref{fig:2}D. For each edge \(e_i\) connecting mating nodes \(m_k\) and \(m_j\), we set a weight to this edge \(f(e_i) = \log(|G_i|)\), where \(|G_i|\) is the total number of possible genotypes of individual \(i\). Essentially, each edge now represents a person.

The problem of finding the optimal set of loop breakers in the original graph is equivalent to removing a set of edges from \(G'\) whose sum of weights is minimal, such that the remaining graph is a tree. This is a well-known problem in graph theory called the MST problem. Prim's algorithm is a widely used algorithm to solve this problem \citep{prim1957shortest}. After applying Prim's algorithm to find the MST of \(G'\), the edges of \(G'\) (individuals) that are not part of the MST form the optimal set of loop breakers. Fig.~\ref{fig:2}E shows the result of finding the MST on Fig.~\ref{fig:2}D given an example set of weights, which implies that individual 8 is selected as the loop breaker in this example.

If at least one person in the trimmed graph is connected to two or more matings (in the trimmed graph) where they are a parent, we refer to this as a “multiple matings” case. Typically, these individuals are also connected to one mating where they are an offspring, but this is not always the case. Regardless, everyone involved in multiple-matings loops will always have trimmed degrees (i.e., degrees in the trimmed graph) greater than or equal to two. Additionally, if someone has a trimmed degree greater than two, the loops they are involved in are guaranteed to be multiple matings loops.

In these instances, the transformation from the trimmed graph to the sub-graph is not guaranteed. A vertex \(v_i\) with a trimmed degree greater than two cannot be represented by a single edge because an edge can only connect two mating vertices, and this person is connected to three or more. Therefore, we often cannot take advantage of Prim’s algorithm to find an optimal set of loop breakers.

Instead, we implemented the simple greedy algorithm discussed by \cite{becker1998automatic} for all multiple-mating cases, which finds close-to-optimal sets of loop breakers in pedigrees with a few loops. Rather than working with a subgraph, this algorithm looks at individuals in the trimmed graph and selects the first one (based on ID) with minimal cost as a loop breaker. Then it begins to break the matings in which this individual participates (as a parent) until no loops remain in the trimmed graph or the loop breaker’s trimmed degree is equal to one. If the pedigree still contains loops and individuals with multiple matings involved in those loops, we repeat the process to select another loop breaker.

To minimize computational complexity \(|G| = \prod_i |G_i|\), or equivalently, \(\log|G| = \sum_i \log|G_i|\), where \(|G_i|\) is the total number of possible genotypes, the set of loop breakers is selected based on two criteria. We choose individuals that have a smaller total number of possible genotypes and/or a higher degree in the trimmed graph. The second criterion is proposed because an individual's trimmed degree serves as an estimate of the number of loops they participate in, and we can break \(l\) loops at one time by adding \(l\) clones of the same individual that participates in all loops. The cost function is defined as \(f(v_i) = \log(|G_i|)/d_i\), where \(d_i\) is the trimmed degree of individual \(i\).The cost function incorporates germline testing data when available. For individuals who have undergone testing for specific genetic variants, their set of possible genotypes \(|G_i|\) is reduced compared to untested individuals. This reduction in \(|G_i|\) directly lowers their cost in the selection algorithm, making tested individuals with definitive results more favorable candidates for loop breaking.

After the greedy algorithm has been called at least once, if loops remain but the remaining individuals do not have multiple matings, we follow the steps in the no-multiple-matings situation to break the remaining loops. Thus, the \texttt{breakloops} algorithm can apply either the greedy or Prim’s algorithm based on the situation, so we call it a hybrid algorithm. The pseudocode for this hybrid algorithm is shown in Algorithm \ref{algo2}.

It is possible for there to be more than one potential loop breaker with the minimal cost in the multiple matings case, or more than one edge with the minimal weight when we apply Prim’s algorithm. In the first case, the first individual encountered with the lowest weight is identified as the loop breaker; in the second, since Prim’s algorithm identifies loop breakers by not selecting them, the last individual with the lowest weight is the chosen loop breaker. For both algorithms, order is based arbitrarily on individual ID numbers.

\begin{algorithm}[H]
\caption{\texttt{breakloops}: Break loops by adding clones of loop breakers}
\begin{algorithmic}[1]
\Require A pedigree $P$
\Ensure Return the initial pedigree if there is no loop, or a modified pedigree with clones of loop breakers if loops exist
\If{$P$ contains no loops}
    \State \Return $P$ \Comment{Return the initial pedigree if there are no loops}
\Else
    \While{loops exist in $P$}
        \State Identify a loop breaker in $P$
        \State Add a clone of the loop breaker to $P$
        \State Modify $P$ by connecting the clone appropriately to break the loop
    \EndWhile
    \State \Return the modified pedigree $P$ \Comment{Return the modified pedigree with loop breakers}
\EndIf
\end{algorithmic}
\label{algo2}
\end{algorithm}

\subsection{Fam3Pro Implementation}
The Fam3Pro package now includes a function to handle pedigrees containing loops through a new Boolean parameter called \textit{breakloop} in the master function. This parameter defaults to FALSE, in which case the software halts with an error message when loops are detected. When set to TRUE, the software invokes the \texttt{breakloops} algorithm to resolve the loops by creating clones, allowing analysis to proceed.

We developed a comprehensive wrapper function named \texttt{breakloops} that coordinates the entire loop-breaking process. This wrapper function requires five essential data columns from the user's pedigree input: "ID," "FatherID," and "MotherID" to determine family structure; "Sex" to properly assign genders when fixing missing parents; and "isProband" to identify individuals of primary interest. Additionally, the function incorporates any available germline testing data to calculate possible genotypes for individuals. In the absence of testing data, the algorithm assigns uniform genotype possibilities to all individuals.

\begin{figure}
    \centering
    \includegraphics[width=1\linewidth]{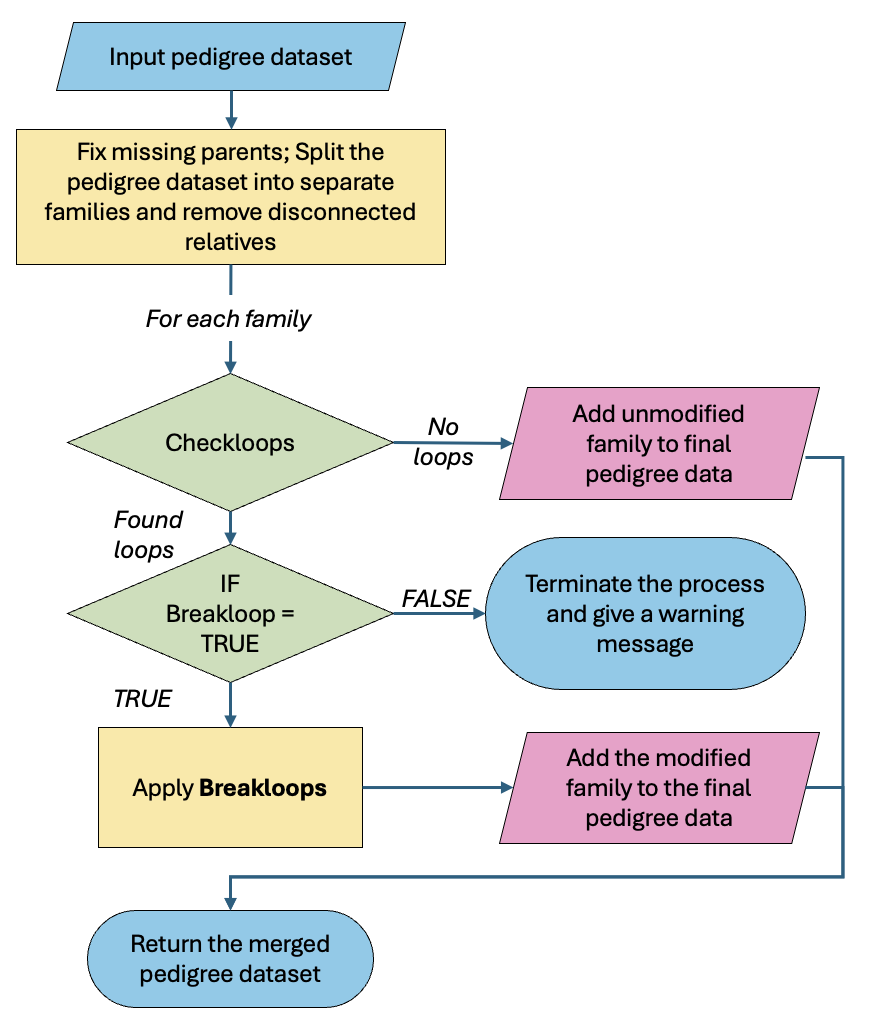}
    \caption{Flowchart of the full implementation of \texttt{breakloops} in Fam3Pro.}
    \label{fig:breakloops_flowchart}
\end{figure}

The \texttt{breakloops} workflow begins with a preparatory phase that addresses common challenges in pedigree data. Figure \ref{fig:breakloops_flowchart} illustrates this process. First, the function addresses incomplete parental information using the \texttt{fixParents} function from the kinship2 package, which creates placeholder parents with no cancer or genetic testing history for individuals with only one reported parent \cite{kinship2}. This ensures that the two-parent requirement for each non-founder is satisfied.

Since the \texttt{breakloops} algorithm operates on connected family units, the function first isolates pedigrees into disjoint subfamilies (if present) and removes individuals unconnected to any proband. This step was originally performed later in the Fam3Pro workflow but was moved earlier to optimize computational efficiency.

After completing these preparatory steps, the wrapper function applies the core \texttt{breakloops} algorithm (Algorithm \ref{algo2}) to each remaining family unit. The algorithm identifies loop breakers, creates appropriate clones, and modifies relationships to eliminate all loops (see Figure \ref{fig:breakloops_flowchart}). The final output is a modified pedigree with no loops and no disconnected individuals, which can then be processed by Fam3Pro's carrier probability and risk estimation functions without encountering computational difficulties from cyclic inheritance patterns.

\section{Results}

To validate the performance and reliability of the \texttt{breakloops} function we employed a two-phase evaluation strategy that encompassed both simulated and real-world clinical data.

We first conducted comprehensive testing of the \texttt{breakloops} function across 17 distinct scenarios to validate its effectiveness in resolving pedigree loops. Our testing methodology involved constructing test pedigrees with predefined loops, creating corresponding expected output pedigrees, applying the \texttt{breakloops} function, and verifying results through direct comparison. The details on the configurations for the tested pedigrees can be found in the Supplementary Material. The \texttt{breakloops} function successfully resolved all loop configurations without exception, demonstrating its robust capability to handle the full spectrum of anticipated pedigree structures.

To validate the function in a clinical context, we applied the updated version of Fam3Pro to a real cohort of 4255 families from Massachusetts General Hospital (MGH) with clinical genetic counseling records. Within this cohort, we identified 12 families with pedigree loops (0.28\%), which aligns with the expected low prevalence of consanguineous relationships in clinical settings.

The identified pedigrees exhibited diverse characteristics, with family sizes ranging from 21 to 47 individuals (mean 31.83). Most families contained a single loop structure, with only one family presenting multiple loops (2 loops). This distribution pattern suggests that while rare, loop structures in clinical pedigrees typically manifest as single, isolated consanguineous relationships rather than complex, interconnected family structures with multiple loops.

Table~\ref{tab:runtime} summarizes the characteristics of these families, including family size and number of loops. The \texttt{breakloops} function successfully resolved all pedigree loops in these real-world examples, demonstrating its practical applicability in clinical genetic counseling settings. The algorithm maintained excellent computational efficiency across all test cases, requiring on average only 0.069 seconds to process each real family with loops (0.08\% of Fam3Pro's total average processing time).

\begin{table}[htbp]
\centering
\caption{Runtime of \texttt{breakloops} in Fam3Pro for 12 Families with Loops}
\begin{tabular}{|c|c|c|c|c|c|}
\hline
\textbf{Family} & \textbf{Size} & \textbf{\#Loops} & \textbf{Fam3Pro (s)} & \textbf{breakloops (s)} & \textbf{\%} \\
\hline
1 & 24 & 1 & 43.01 & 0.053 & 0.12 \\
2 & 33 & 1 & 101.95 & 0.063 & 0.06 \\
3 & 28 & 1 & 85.69 & 0.060 & 0.07 \\
4 & 21 & 1 & 37.43 & 0.050 & 0.13 \\
5 & 41 & 1 & 124.96 & 0.069 & 0.06 \\
6 & 47 & 2 & 167.53 & 0.084 & 0.05 \\
7 & 40 & 1 & 107.34 & 0.073 & 0.07 \\
8 & 22 & 1 & 54.74 & 0.053 & 0.10 \\
9 & 33 & 1 & 73.96 & 0.062 & 0.08 \\
10 & 30 & 1 & 76.41 & 0.061 & 0.08 \\
11 & 31 & 1 & 74.31 & 0.100 & 0.13 \\
12 & 32 & 1 & 40.54 & 0.101 & 0.25 \\
\hline
\textbf{Average} & \textbf{31.83} & \textbf{1.08} & \textbf{82.32} & \textbf{0.069} & \textbf{0.08} \\
\hline
\end{tabular}
\label{tab:runtime}
\end{table}

\section{Discussion}

In this paper, we present a new feature for Fam3Pro to handle pedigree loops called \texttt{breakloops}. This algorithm addresses a significant limitation in previous versions of Fam3Pro by enabling analysis of family structures with consanguineous relationships. The hybrid approach we implemented is optimal for pedigrees without individuals with multiple matings in the loops and close-to-optimal otherwise. To ensure compatibility with Fam3Pro's underlying architecture, the algorithm includes preprocessing steps to fix missing parents for non-founders, partition pedigrees into separate family units, and remove individuals disconnected from probands.

Our \texttt{breakloops} algorithm builds upon the foundation established by Becker and Geiger's LoopBreaker algorithm \cite{becker1998automatic}. Similar to their approach, we route pedigrees to different sub-algorithms based on the presence or absence of multiple matings, employing the same greedy algorithm for multiple matings scenarios. For loops without multiple matings, however, we diverge in implementation details. While \cite{becker1998automatic} apply Kruskal's algorithm directly to the trimmed graph, we utilize Prim's algorithm on a derived subgraph. A key practical advantage of our implementation is that it automatically identifies optimal loop breakers without requiring user input, whereas LoopBreaker requires users to manually specify valid loop breakers.

The validation approach included testing on simulated families as well as using real-world clinical data. The \texttt{breakloops} function successfully resolved all loop configurations, demonstrating its robustness and applicability to actual clinical scenarios.

Although the algorithm effectively resolves a wide variety of pedigree loops, certain edge cases may not achieve optimal resolution, particularly in scenarios involving founders in loops and multiple-matings situations. Future enhancements could focus on refining the decision criteria to enable optimal breaking for these cases and ensuring clones maintain identical properties to their source individuals throughout the analysis workflow.

In conclusion, the addition of \texttt{breakloops} significantly extends Fam3Pro's capabilities, enabling risk assessment for a broader range of family structures. This enhancement allows genetic counselors and clinicians to provide more inclusive services to patients whose family histories include consanguineous relationships. By addressing both the theoretical challenges of loop detection and the practical needs of clinical genetic counseling, this feature represents an important advancement in hereditary cancer risk assessment software. The \texttt{breakloops} feature is available in Fam3Pro version 2.0.0, making these capabilities accessible to the clinical and research communities.

\bibliographystyle{plainnat}
\bibliography{refs/BayesMendelBibliography, refs/GPBibliography}

\clearpage
\appendix

\setcounter{figure}{0}
\setcounter{table}{0}
\setcounter{equation}{0}

\renewcommand{\thefigure}{A\arabic{figure}}  
\renewcommand{\thetable}{A\arabic{table}}    
\section{Supplementary Material: Comprehensive Testing of the Breakloops Function}

To validate the \texttt{breakloops} functionality across diverse scenarios, we conducted systematic testing using 17 distinct test cases encompassing various loop configurations. The \texttt{breakloops} function successfully identified and resolved all loops across all test scenarios. This supplement describes our testing methodology and the specific scenarios examined.

For each test case, we employed the following protocol:
\begin{itemize}
    \item We constructed a test pedigree containing one or more predefined loops.
    \item We created a corresponding ``clones'' pedigree representing the expected output after proper loop resolution.
    \item We applied the \texttt{breakloops} function to the test pedigree.
    \item We verified correctness by comparing the function's output to the expected pedigree using the \texttt{identical()} function.
\end{itemize}

The test cases covered various loop sizes (3 to 7 individuals), the presence or absence of individuals with multiple matings, scenarios with multiple simultaneous loops, and edge cases such as missing parental information. For example, test cases T1–T7 include loops without multiple matings ("No-MM"), while T8–T14 feature loops where at least one individual has multiple matings ("MM"). Test cases T15 and T16 combine both types within a single pedigree. T17 tests the function’s robustness when handling incomplete parental data. Table~\ref{tab:scenarios} provides a detailed summary of all test scenarios.

\begin{table}[htbp]
\centering
\small
\caption{Summary of Testing Scenarios}
\setlength{\tabcolsep}{4pt}
\begin{tabular}{c|c|c}
\hline
\textbf{ID} & \textbf{Scenario} & \textbf{Description} \\
\hline
T1 & Small Loop & Loop consisting of 3 individuals \\
T2 & Medium Loop & Loop consisting of 4 individuals \\
T3 & Large Loop & Loop consisting of 5 individuals \\
T4 & XL Loop & Loop consisting of 7 individuals \\
T5 & Multiple Loops & Scenario with 3 different loops \\
T6 & Same Weights & All individuals have same weight \\
T7 & Proband Ideal & Proband is ideal loop breaker \\
T8 & Small Loop (MM) & Loop with multiple matings case, 3 individuals \\
T9 & Medium Loop (MM) & Loop with multiple matings case, 4 individuals \\
T10 & Large Loop (MM) & Loop with multiple matings case, 5 individuals \\
T11 & XL Loop (MM) & Loop with multiple matings case, 7 individuals \\
T12 & Multiple Loops (MM) & 3 different loops, some with multiple matings \\
T13 & Same Weights (MM) & Equal weights, multiple matings included \\
T14 & Proband Ideal (MM) & Proband is ideal breaker, includes multiple matings \\
T15 & Mixed Loops 1 & One MM loop and two No-MM loops \\
T16 & Mixed Loops 2 & Two MM loops and one No-MM loop \\
T17 & Missing Parent & One or more parents missing \\
\hline
\end{tabular}
\label{tab:scenarios}
\end{table}

In conclusion, our comprehensive testing demonstrates that the \texttt{breakloops} function robustly handles a wide spectrum of pedigree configurations and edge cases. The successful resolution of all test scenarios confirms that the algorithm can effectively identify optimal loop breakers, create appropriate clones, and reconstruct pedigrees without loops while preserving the essential genetic relationships. 

\end{document}